\documentclass[useAMS,usenatbib]{mn2e}
\input psfig.sty
\usepackage[dvips]{color}
\usepackage{graphicx}
\usepackage{epsfig}

\title[GRB Hubble diagram and constraints on a $\Lambda$(t)CDM model]
  {GRB Hubble diagram and constraints on a $\Lambda$(t)CDM model}
\author[H. Velten, A. Montiel and S. Carneiro]{H. Velten$^{1}$\thanks{E-mail:
velten@physik.uni-bielefeld.de}, A. Montiel$^{2}$\thanks{E-mail: amontiel@fis.cinvestav.mx} and S. Carneiro$^3$\thanks{E-mail: saulo.carneiro.ufba@gmail.com}\\
$^1$Fakult\"at f\"ur Physik, Universit\"at Bielefeld, Postfach 100131, 33501 Bielefeld, Germany\\$^2$Dpto. de F\'isica, Centro de Investigaci\'on y de Estudios Avanzados del I. P. N.,
Av. IPN 2508, D.F., Mexico\\$^3$Instituto de F\'{\i}sica, Universidade Federal da Bahia, 40210-340, Salvador, BA, Brazil}
\begin{document}
\date{\today}

\pagerange{\pageref{firstpage}--\pageref{lastpage}} 

\def\LaTeX{L\kern-.36em\raise.3ex\hbox{a}\kern-.15em
    T\kern-.1667em\lower.7ex\hbox{E}\kern-.125emX}
\label{firstpage}

\maketitle

\begin{abstract}
In previous papers, a cosmological model with constant-rate particle creation and vacuum term decaying linearly with the Hubble parameter was shown to lead to a good concordance when tested against precise observations: the position of the first peak in the spectrum of anisotropies of the cosmic microwave background (CMB), the Hubble diagram for supernovas of type Ia (SNe Ia), the distribution of large-scale structures (LSS) and the distance to the baryonic acoustic oscillations (BAO). That model has the same number of parameters as the spatially flat standard model and seems to alleviate some observational/theoretical tensions appearing in the later. In this letter we complement those tests with $109$ gamma ray bursters (GRB), $59$ of them with redshifts above $z=1.4$, which permits to extend the Hubble diagram to redshifts up to $z\approx 8$. For the calibration of the $50$ GRBs with $z<1.4$ we use the $288$ supernovas of the Sloan Digital Sky Survey (SDSS) project, calibrated with the MLCS2k2 fitter, less model-dependent than other samples like Union2. Our results show a good concordance with the previous tests and, again, less tensions between SNe Ia and GRB best fits as compared to the standard model.
\end{abstract}


\section{Introduction}

In the last decade an unprecedented amount of precise observations has led to the consolidation of the so called standard cosmological model, with a geometry given by a spatially flat Friedmann-Lema\^{\i}tre-Robertson-Walker (FLRW) metric and an energy content formed by around $5\%$ of baryons, $23\%$ of cold dark matter and $72\%$ associated to a cosmological constant $\Lambda$, the component responsible for the present acceleration of the expansion \citep{ParticleDataGroup}. The $\Lambda$CDM model, together with the standard description of the early hot phase of universe's evolution and the inflationary paradigm for the generation of primordial density fluctuations, has been shown robust when tested against the more precise observations: i) the spectrum of anisotropies of the CMB temperature \citep{komatsu}; ii) the mass power spectrum of LSS distribution \citep{montesano}; iii) the Hubble diagram for supernovas of type Ia \citep{SNUNION,hicken,SNSNLS}; and (in a still lower precision level) the distance scale to the BAO peak \citep{wiggleZ}. The two later are background tests of the redshift-distance relation predicted by the model, and involve relatively low redshifts (up to $z\approx 2$). The two former also test predictions at the perturbative level and up to the redshift of last scattering and beyond. For example, the matter density at the time of matter-radiation equality (around $z\approx 3300$ for the $\Lambda$CDM model) is essential to determine the turn-over of the mass power spectrum and hence its present shape. In the case of the CMB anisotropy spectrum, many free parameters are available in the data fitting. If we are just interested in fixing background parameters like the present relative matter density for example, a simplified test is given by fitting the position of the first acoustic peak in the spectrum.

In spite of the relative success of the standard model in getting a good concordance between those observations, some theoretical and observational problems have stimulated the study of alternative models of dark energy. Among the theoretical problems usually mentioned we can cite the absence of a microphysical explanation for the tiny value of the cosmological constant and the approximate coincidence between $\Lambda$ and the present matter density. On the observational level, the celebrated concordance is actually dependent on the sample of SNe Ia used in the joint analysis \citep{tension,tension2}. Among the recent compilations, the Union2 \citep{SNUNION} is one of the biggest samples and, alone, predicts for the $\Lambda$CDM a present matter density parameter $\Omega_{m0} \approx 0.27$. However, we should have in mind that the Union2 dataset is calibrated with the Salt2 fitter, which makes use of a fiducial $\Lambda$CDM model for including high-z supernovas in the calibration. Therefore, that sample is not model-independent and the test should be viewed as rather a test of consistence.

Model-independent SNe samples can be obtained by using instead the MLCS2k2 fitter, which uses only low and intermediate redshifts in the calibration \citep{MLCS2K2}. Although this may lead to other kinds of limitations, it is a must if we want to test the robustness of the concordance. By using, for instance, the Constitution and SDSS samples calibrated with that fitter, we obtain $\Omega_{m0} \approx 0.33$ and $\Omega_{m0} \approx 0.40$, respectively (see e.g. Table 2 of reference \citep{jailson2011}), values in clear tension with that obtained from LSS. For example, the best-fit value obtained from the analysis of the 2dFGRS data for the mass power spectrum is $\Omega_{m0} \approx 0.23$ \citep{2dFGRS}. A similar result, $\Omega_{m0} \approx 0.22$, is obtained from the analysis of the shape of the SDSS DR5 galaxy power spectrum when only large scales ($0.01<k<0.06$ $h$Mpc$^{-1}$) are included \citep{Percival}. This tension is also clear when we look the reduced $\chi^2$'s of the joint analysis including the four tests referred above (see Table 1 below). 

Such a tension may be just the result of systematics in the samples or in the calibration. But it may also have some physics behind it. As we have commented above, the mass power spectrum is particularly sensitive to the matter density at the time of matter-radiation equality. While the SNe Ia distance relations are tested for relatively low redshifts. Therefore, the mentioned tension may be suggesting a late-time process of particle creation from the vacuum, in such a way that we have at early times the matter density needed to fit the LSS observations, but a present density above that expected if no matter has been created.

In order to verify this hypothesis, the data must be fitted by a model with particle creation. Since the radiation epoch must remain unaffected, no photons should be produced. In the same way, production of baryons is not allowed if we want to maintain the main features of CMB. Therefore, only dark particles will be created in this model. Furthermore, since the observed dark matter is cold, the created matter is pressureless. The simpler way of realizing this phenomenological recipe is with a constant rate of creation. As we will see, the corresponding model is uniquely defined, has the same number of free parameters as the standard model and is not reducible to the later. The data analysis with this model leads to a good concordance. With the Union2 sample the fit is as good as in the standard case, while the use of the Constitution and SDSS samples leads to a very good agreement with the value derived from the LSS analysis in this model, $\Omega_{m0} \approx 0.45$. We will also see that the backreaction associated to the matter creation implies a time-dependent vacuum density, linearly proportional to the Hubble parameter, in order to respect the conservation of total energy.

In addition to the main observational tests referred above, several others can be done, despite the less precision involved. Among them we can cite, for example, weak gravitational lensing, gas mass fraction in clusters, redshift-distance relations for compact x-ray sources, the redshift-age relation for high-$z$ objects, and the peculiar velocity field of galaxies. For elucidating the problem we are discussing here, i.e., the tension between early and late-time predictions for the matter density, an interesting complementary test is provided by gamma ray bursts (GRB) \citep{GRB1}, since they can extend the Hubble diagram up to redshifts $z \approx 8$ \citep{Schaefer:2007,WH}. Actually, the use of GRBs as standard candles is quite controversial, as we will discuss in the Section III. In this paper we do assume that GRBs can be used as cosmological probes, following references \citep{Cuesta,GRB2,GRB3,Velten,MB}. Once more, any strong model-dependence should be avoided and, for this goal, new GRB samples will be generated in this letter, calibrated with help of the SDSS (MLCS2k2) SNe Ia compilation discussed above.

\section{The model}

Consider the possibility of creation of non-relativistic dark particles at a constant rate $\Gamma$. The balance equation for this process (some times called the Boltzmann equation) is given by
\begin{equation} \label{Boltzman}
\frac{1}{a^3}\frac{d}{dt}\left(a^3n\right)= \Gamma n,
\end{equation}
where $n$ is the particle number density at a given time. By multiplying by the created particle mass $M$, it can be rewritten as
\begin{equation} \label{conservacao}
\dot{\rho}_m + 3H\rho_m = \Gamma \rho_m,
\end{equation}
where $\rho_m = nM$  is the matter density.

In addition, let us take the Friedmann equation for the spatially flat case,
\begin{equation} \label{Friedmann}
\rho = \rho_m + \Lambda = 3H^2,
\end{equation}
where $\rho_m$ is the matter density and $\Lambda$ is the energy density associated to a vacuum term. Combining the two above equations and using for the vacuum term the equation of state $p_{\Lambda} = - \Lambda$, we can obtain
\begin{equation} \label{conservacao2}
\dot{\rho} + 3H(\rho +p) = 0
\end{equation}
(where $\rho$ and $p$ refer to the total density and pressure, respectively), provided we take $\Lambda = 2\Gamma H + \lambda_0$, where $\lambda_0$ is a constant of integration. Since there is no microphysical scale defining such a constant, we will take $\lambda_0 = 0$. Therefore, we have\footnote{Strictly speaking, this result is only exact if we dismiss the conserved baryons in the balance equations. Since baryons represent only about $5\%$ of the total energy content, relation (\ref{linear}) can be considered a good approximation.}
\begin{equation} \label{linear}
\Lambda = 2\Gamma H,
\end{equation}
and Eq. (\ref{conservacao2}) can be rewritten as
\begin{equation} \label{conservacao3}
\dot{\rho}_m + 3H\rho_m = -\dot{\Lambda}.
\end{equation}
We can see that the creation of pressureless matter is only possible with a concomitant time-variation of the vacuum density, which acts as a source of particles in the conservation equation, Eq.  (\ref{conservacao3}), decaying linearly with the Hubble parameter $H$.

Dividing Eq. (\ref{linear}) by $3H^2$ and using $\Lambda/(3H^2) \equiv 1 - \Omega_m$, we obtain
\begin{equation} \label{rate}
\Gamma = \frac{3}{2} \left( 1 - \Omega_{m} \right) H,
\end{equation}
which is valid for any (late) time. For the present time (with $\Omega_{m0} \approx 1/3$) we have $\Gamma \approx H_0$. On the other hand, in the future de Sitter limit ($\Omega_m \rightarrow 0$), it follows that $\Gamma \approx (3/2)H_{dS}$. That is, apart from a constant factor, the creation rate is equal to the Gibbons-Hawking temperature associated to the de Sitter horizon \citep{Gibbons,Gibbons2}.

With $\Lambda = 2\Gamma H$ it is easy to obtain the Hubble function \citep{Borges,jailson2006}
\begin{equation}\label{Hgeral}
\frac{H}{H_0} = 1 - \Omega_{m0} + \Omega_{m0} (1
+ z)^{3/2},
\end{equation}
where $\Omega_{m0}$ and $H_0$ (the present values of the relative matter density and of the Hubble parameter, respectively) are, like in the standard model, the only free parameters to be adjusted. In the asymptotic limit $z \rightarrow -1$ it leads to a constant $H$, that is, to a de Sitter universe. On the other hand, for early times (high $z$), we have $H\approx H_0 \Omega_{m0} z^{3/2}$, which, through the Friedmann equation, leads to $\rho_m = 3H_0^2\Omega_{m0}^2 z^3$. Therefore, at early times the energy density scales with $z^3$, that is, with $a^{-3}$, as in the standard model. However, in the later we would have $\rho_m = 3H_0^2\Omega_{m0}z^3$. The extra factor $\Omega_{m0}$ ($< 1$) in the former is related to the late-time matter creation. In order to have the same amount of matter today, we should have less matter at higher redshifts. Reversely, if we want the same amount of matter at high $z$ (in order to preserve the observed features of CMB and LSS), we expect a present matter density higher than in the standard model.

This is confirmed when we perform precise observational tests with the present model, at both the background and perturbative levels. In previous publications \citep{Borges,jailson2006,jailson2008,jailson2011,Fabris,Zimdahl,PLB} the model has been tested against observations of the Hubble diagram for type Ia supernovas, the position of the first peak in the CMB spectrum of anisotropies, the power spectrum of large scale structures and the distance to baryon acoustic oscillations. The concordance is quite good and, depending on the sample of supernovas used, it is actually better than in the $\Lambda$CDM case, as shown in Table 1 (with $H_0$ marginalized).

\begin{table*}
\begin{center}
\caption{Limits to $\Omega_{m0}$ (SNe+ CMB + BAO+LSS) \citep{PLB}.}
\begin{tabular}{r@{~~}c@{~~}c@{~~}c@{~~}c@{~~}}
\hline \hline \\
\multicolumn{1}{c}{ } & \multicolumn{2}{c}{$\Lambda(t)$CDM } & \multicolumn{2}{c}{$\Lambda$CDM } \\
\multicolumn{1}{c}{Test}&
\multicolumn{1}{c}{$\Omega_{m0}$\footnote{Error bars stand for $2\sigma$}}&
\multicolumn{1}{c}{$\chi^2_{min}/\nu$}&
\multicolumn{1}{c}{$\Omega_{m0}$ $^a$}&
\multicolumn{1}{c}{$\chi^2_{min}/\nu$}\\ \hline \\
Union2 (SALT2)......&$0.420^{+0.009}_{-0.010}$ & 1.063 & $0.235\pm 0.011$ & 1.027 \\
SDSS (MLCS2k2).......& $0.450^{+0.014}_{-0.010}$ & 0.842 & $0.260^{+0.013}_{-0.016}$ & 1.231 \\
Constitution (MLCS2k2[17]).......& $0.450^{+0.008}_{-0.014} $ &1.057 & $0.270\pm 0.013$ & 1.384\\
\hline \hline
\end{tabular}
\end{center}
\end{table*}

We can see in the table that, when the Union2 compilation is used in the joint analysis, similar reduced $\chi^2$'s are obtained for both models. However, Union2 data are calibrated with the Salt2 fitter by making use of a fiducial $\Lambda$CDM model in the calibration process. A model-independent analysis is possible if we consider instead the MLCS2k2 fitter in the calibration, as is the case of the SDSS and Constitution samples in Table 1. With these compilations we obtain a better concordance with the present model. As anticipated above, the present matter density is higher than in the standard case, with $\Omega_{m0} \approx 0.45$.

\section{The GRB samples}

The luminosity of GRBs appears to be correlated with their temporal and spectral properties and, although these correlations are not yet fully understood from first principles, their existence has naturally suggested the use of GRBs as distance indicators, offering a possible route to probe the expansion history of the Universe up to redshifts $z \approx 8$ \citep{WH}. 

The typical spectrum of the prompt emission of GRBs can be expressed as an exponentially connected broken power-law, the so-called Band function \citep{Band:1993}. Then we can determine the spectral peak energy $E_p$, the photon energy at which
the $\nu F_{\nu}$ is brightest.

The two most used correlations are the Amati relation $E_{p}-E_{iso}$, relating the rest frame energy of the spectra $E_{p}$ and the emitted isotropic energy $E_{iso}$, and the Ghirlanda  relation $E_{\gamma}-E_{p}$, relating $E_{p}$ with $E_{\gamma}=E_{iso} (1 - \cos{\theta_{jet}})$. The latter correlation is the tightest of the GRBs calibration relations. However, to be included in this relation the GRB afterglow must have an observed jet break in its light curve, thus only a fraction of the observed events can contribute for establishing this relation. Therefore, the Amati relation is more statistically reliable as it can be calibrated with a greater number of events.

We shall apply here Amati's power law expression \citep{Amati:2002,Amati:2008}, relating $E_p=E_{p, obs}(1+z)$ with the isotropic equivalent radiated energy $E_{iso}$, defined as
\begin{equation}
E_{iso}=4\pi d^2_L S_{bolo} (1+z)^{-1},
\label{Eq:Eiso}
\end{equation}
where $S_{bolo}$ is the bolometric fluence of gamma rays in the GRB at a redshift $z$ and $d_L$ is the luminosity distance to it. Similarly to \citep{Schaefer:2007}, we will rewrite the Amati relation as
\begin{equation}
\log \frac{E_{iso}}{\mathrm{erg}}=\lambda + b\log \frac{E_{p}}{300\mathrm{keV}}.
\label{Eq:Amati}
\end{equation}
 
GRBs can be considered as distances indicators if they can be calibrated at low redshifts. However, a few data are available at low redshifts. To mend this lack of data most calibrations take for granted a particular cosmological model. In order to avoid this circularity problem here we follow the method proposed in \citep{Liang:2008}, which consists in using SNe Ia as a distance ladder to calibrate the GRBs. Note that with the well-known relation 
\begin{equation}
 \mu= 5\log \frac{d_L}{\mathrm{Mpc}}+25,
\label{Eq:mu}
\end{equation}
one can convert the distance moduli $\mu$ into luminosity distances $d_L$ (in units of Mpc).

Since the distance moduli for the SNe Ia are known, a cubic interpolation is performed to determine the parameters $\lambda$ and $b$ in the Amati relation for the low-redshift GRBs ($z < 1.4$). Then, the distance moduli for the high-$z$ GRBs are obtained and they can be used as standard candles without the circularity problem. It is worth mentioning that the calibration of GRBs is still a quite controversial subject. Indeed, even the Amati relation has been contested by some authors (see, for instance, \citep{Collazzi:2011}, where it is argued that the Amati relation could be an artifact of selection effects in both the burst population and the detector). With this cautionary remark, we proceed with the calibration.

We perform the cubic interpolation by two methods: splines and Hermite. A spline is a polynomial between each pair of table points, but one whose coefficients are determined ``slightly" nonlocally. The nonlocality is designed to guarantee global smoothness in the interpolated function up to some order of derivatives. The cubic spline interpolation produces an interpolated function that is continuous through the second derivative \citep{NumericalRecipes}. On the other side, Hermite interpolation is also a method of interpolating data-points as a polynomial function. The generated Hermite polynomial is closely related to the Newton polynomial, in that both are derived from the calculation of divided differences. Hermite interpolation (first-derivative continuous) requires four points so that it can achieve a higher degree of continuity \citep{Stoer:1993}.

So, by using at low redshifts the $288$ SDSS SNe Ia data \citep{Kessler:2009}, we derived the distance moduli for the $50$ low-redshift ($z<1.4$) GRBs by cubic-splines and cubic-Hermite interpolations. From Eq. (\ref{Eq:Eiso}) with the corresponding $S_{bolo}$ and luminosity distances $d_L$, we then derived $E_{iso}$ for these $50$ GRBs. Furthermore, with the corresponding $E_{p}$ we found the best fit for the Amati relation, Eq.  (\ref{Eq:Amati}): 
\begin{equation}
\lambda=52.8076 \pm 0.0683, \quad b=1.6909 \pm 0.1156,
\label{Eq:CubicSplines}
\end{equation}
for the cubic-splines interpolation, and
\begin{equation}
\lambda=52.8960 \pm 0.0874, \quad b=1.7851 \pm 0.1480,
\label{Eq:CubicHermite}
\end{equation}
for the cubic-Hermite one.
Extrapolating these calibrated Amati relations to the $59$ high-redshift ($z>1.4$) GRBs given in Table 2 of \citep{WH}, the distance moduli $\mu$ are obtained. Only these new 59 datapoints are then used in order to test the cosmological model \footnote{Both samples are available under request.}.

\section{Results and Discussion}

In the last section, assuming the SDSS (MLCS2k2) SNe Ia sample as our calibration source at low redshifts, we have produced two new GRB data sets based on the Hermite and the spline interpolation methods. Before proceeding with a statistical analysis with our new samples we consider three previous GRB data sets: i) the NPZ sample \citep{Liang:2011} with $27$ GRBs at $z<1.4$ and the remaining $42$ objects in the range $1.4<z<6.6$; ii) the Hymnium sample \citep{WH}, with $50$ low-redshift data points and $59$ GRBs in the range $1.4<z<8.2$; iii) the MB sample \citep{MB} that makes use of the same GRB compilation as the Hymnium data, but using a different interpolation for the $50$ low-redshift GRBs, leading to a different extrapolated Hubble diagram at high $z$. These samples share the same SNe Ia calibrating source, the Union2 data. A direct comparison of the $\Lambda$(t)CDM model with such samples retains information from the Union2 data set and, consequently, an imprint of the $\Lambda$CDM model. Nevertheless, for the sake of comparison, we will perform a statistical analysis with all the above GRB samples. 

Assuming a set of free parameters ($\Omega_{m0}, h$), where $H_0=100h~$Km s$^{-1}$Mpc$^{-1}$, the agreement between theory and observation is measured by minimizing the $\chi^2$,
\begin{equation}
\chi^2\equiv\chi^2(\Omega_{m0}, h)=\sum^{N}_{i=1}\frac{\left(\mu^{th}_i-\mu^{obs}_i\right)^2}{\sigma^2_i},
\end{equation}
where $\mu^{th}$ and $\mu^{obs}$ are the theoretical and observed values of the distance moduli, respectively, and $\sigma$ means the error for each one of the $N$ data points. The $68.3\%$, $95.4\%$ and $99.73\%$ confidence levels (C.L.) are determined by $\Delta\chi^2\equiv \chi^2 - \chi^2_{min} \leq 2.3$, $6.17$ and $11.8$, respectively.

\begin{figure*}
\vspace{.2in}
\centerline{\psfig{figure=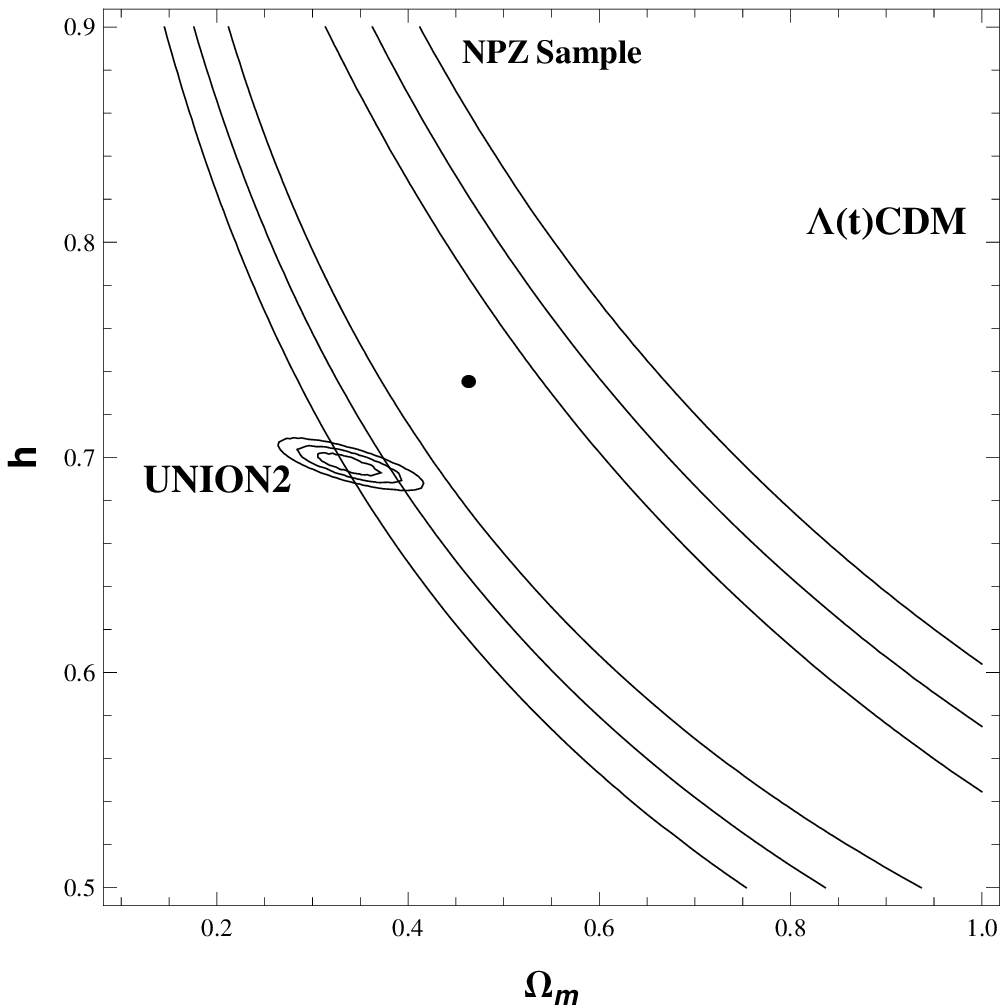,width=2.2truein,height=2.2truein}
\psfig{figure=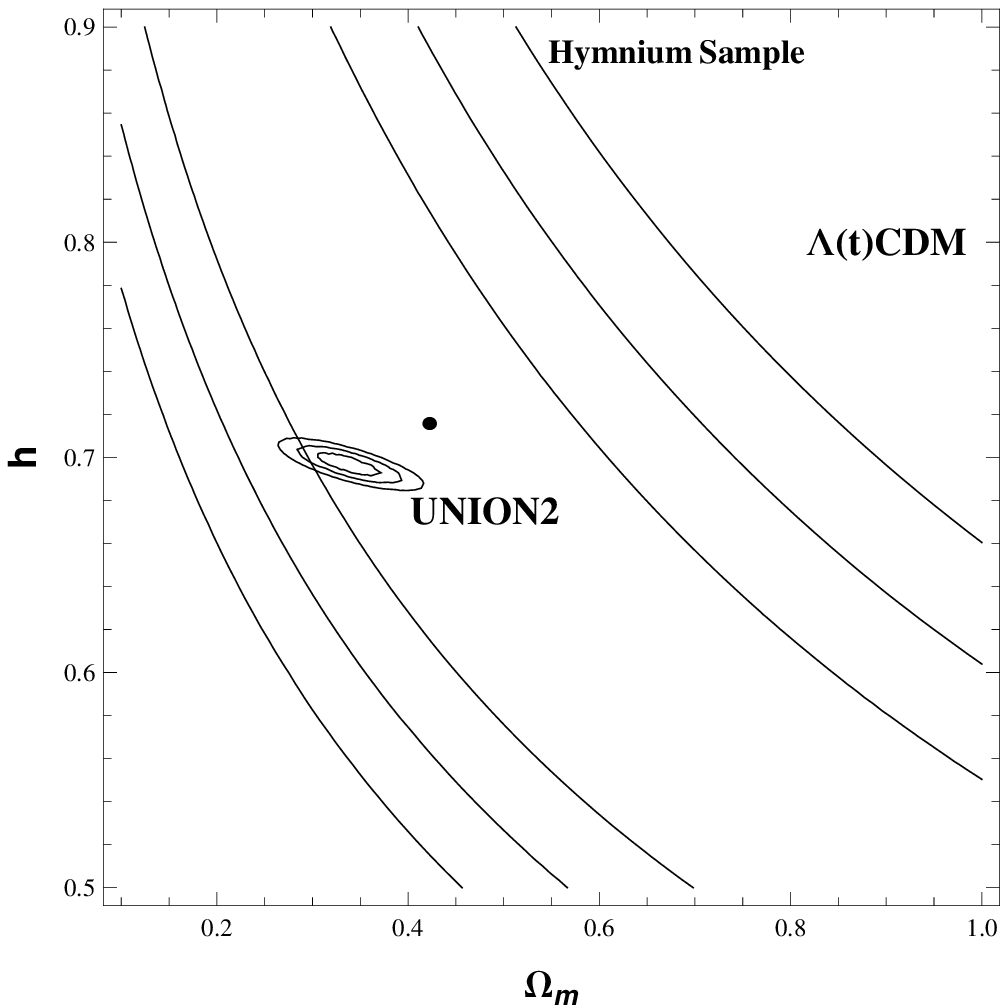,width=2.2truein,height=2.2truein}
\psfig{figure=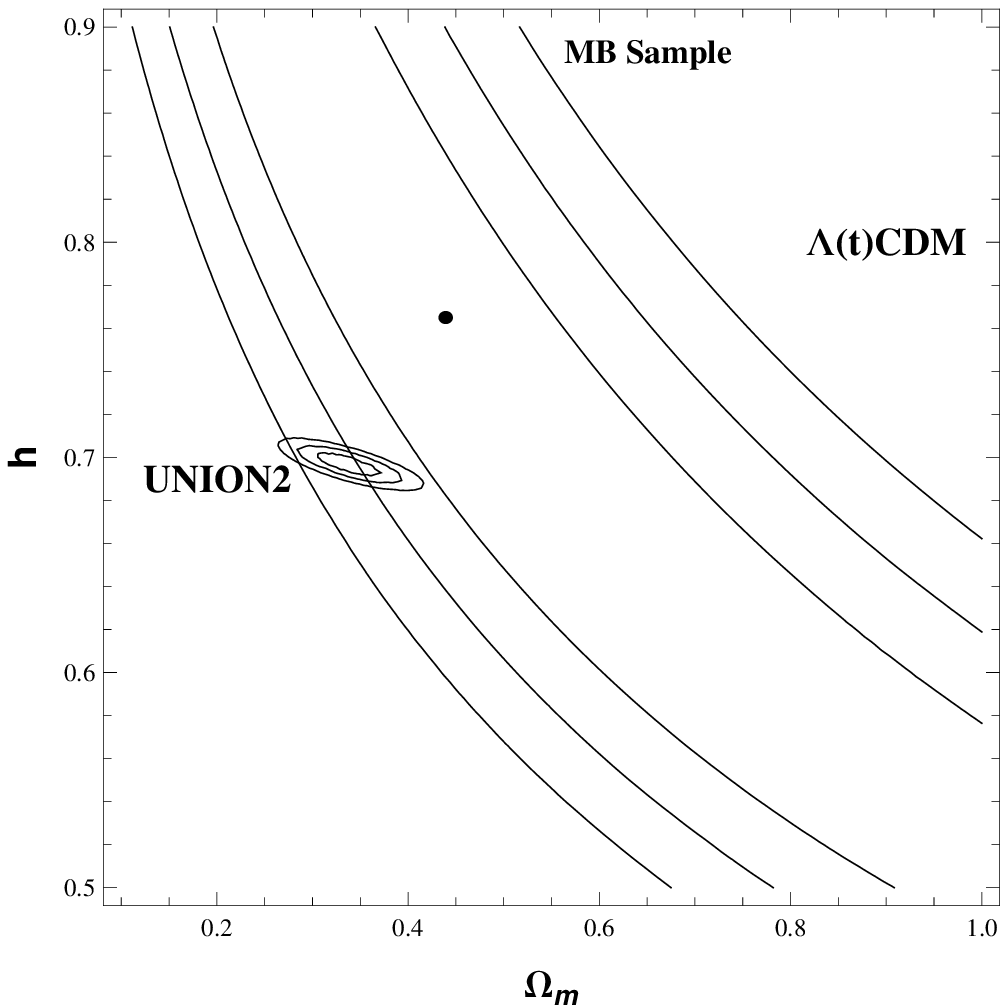,width=2.2truein,height=2.2truein}}
\caption{Constraints on the free parameters of the $\Lambda$(t)CDM model. The black dot is the best fit for each GRB sample. All GRB samples have been calibrated with the Union2 SNe Ia data set (closed elliptical contours).}
\label{fig1}
\end{figure*}

\begin{figure*}
\vspace{.2in}
\centerline{\psfig{figure=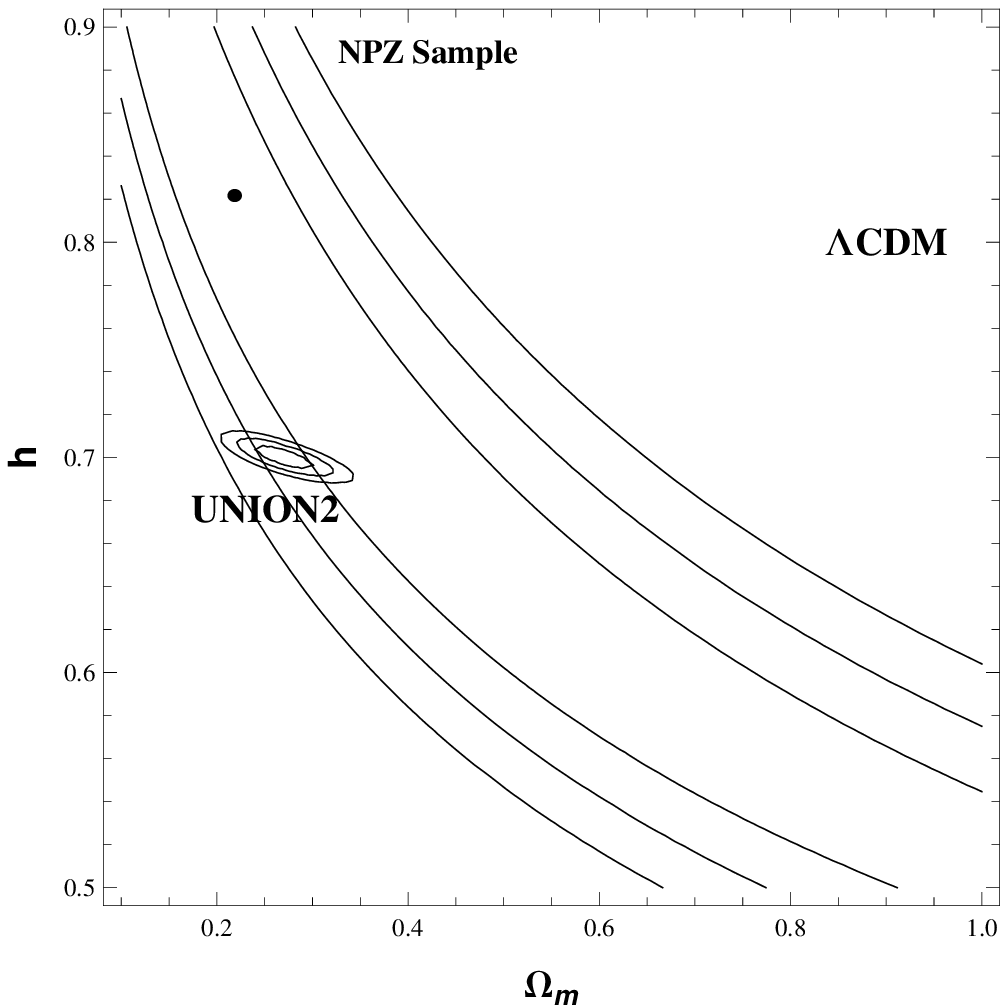,width=2.2truein,height=2.2truein}
\psfig{figure=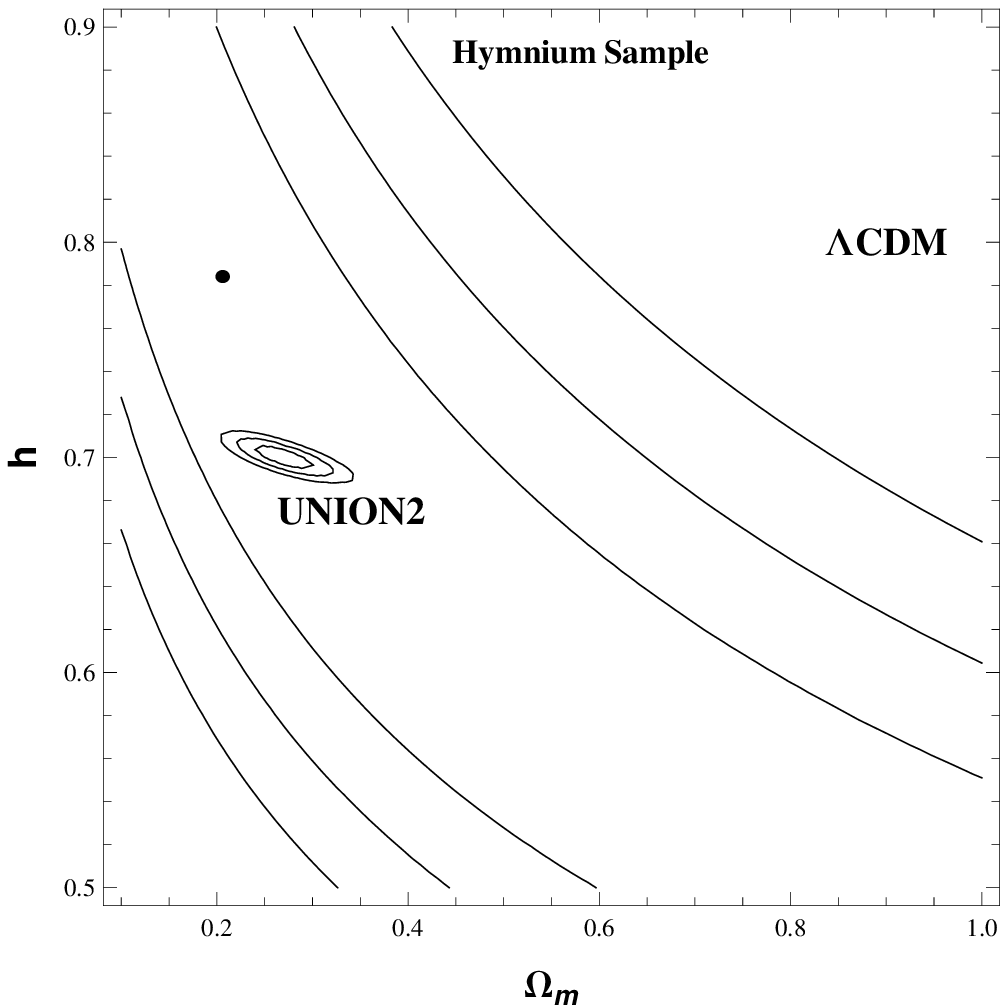,width=2.2truein,height=2.2truein}
\psfig{figure=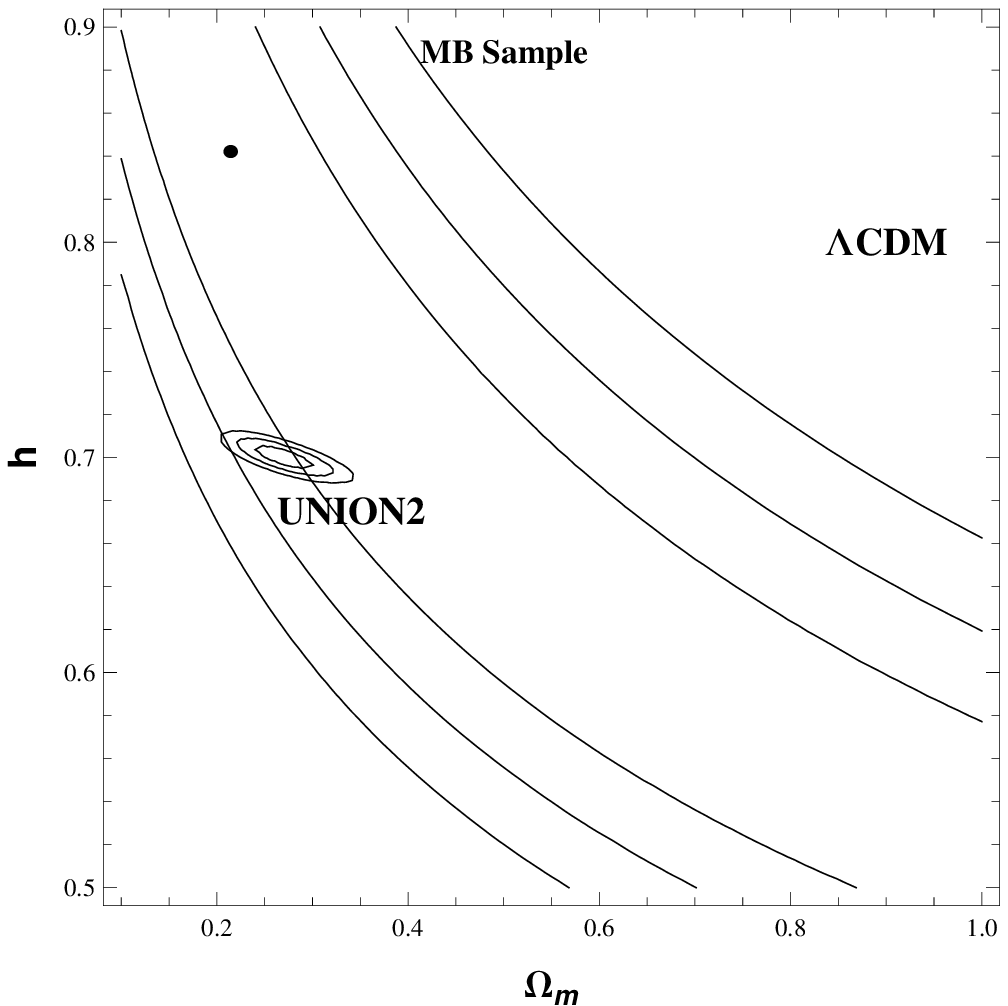,width=2.2truein,height=2.2truein}}
\caption{Constraints on the free parameters of the $\Lambda$CDM model. The black dot is the best fit for each GRB sample. All GRB samples have been calibrated with the Union2 SNe Ia data set (closed elliptical contours).}
\label{fig2}
\end{figure*}

Fig. \ref{fig1} shows the constraints on the free parameters ($\Omega_{m0},h$) for the $\Lambda$(t)CDM model using the above mentioned NPZ (left panel), Hymnium (central panel) and MB (right panel) samples. The contours for the Union2 SNe Ia sample are also shown. In spite of the large dispersion observed in the GRB samples one can infer from their best fits (NPZ: ($0.463, 0.727$), Hymnium: ($0.422, 0.708$), MB: ($0.439, 0.757$)) that they agree with larger values for the amount of dark matter and the Hubble constant as compared with the Union2 best fit ($\Omega_{m0}=0.337, h=0.697$). A similar analysis for the $\Lambda$CDM model is displayed in Fig. \ref{fig2}. Note that in this case very large values for the Hubble constant are predicted. We summarize the results of Figs. \ref{fig1} and \ref{fig2} in Table 2.

\begin{figure*}
\vspace{.2in}
\centerline{
\psfig{figure=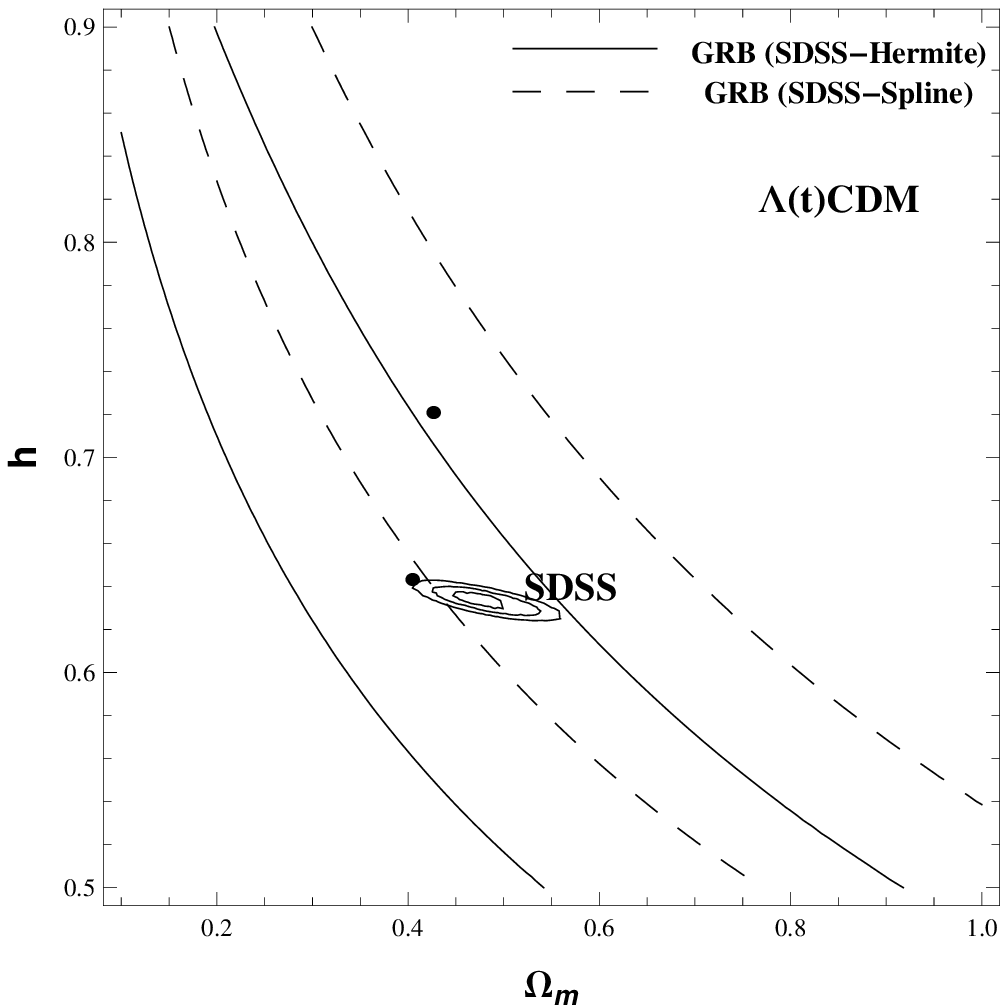,width=2.9truein,height=2.75truein}\hspace{.1in}
\psfig{figure=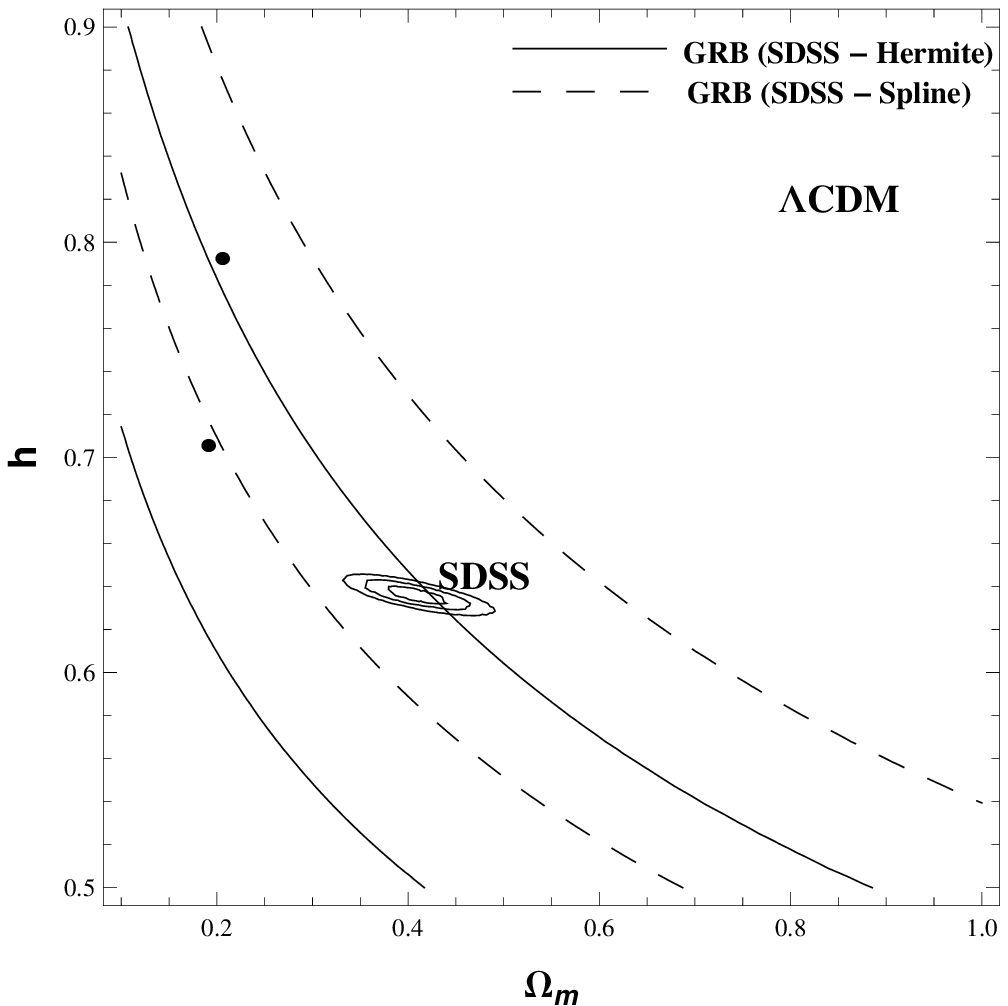,width=2.9truein,height=2.75truein}}
\caption{Contraints on the free parameters of the $\Lambda$(t)CDM ($\Lambda$CDM) model (left (right) panel). The $59$ GRBs at $z>1.4$ were calibrated with the SDSS (MLCS2k2) SNe Ia data set using cubic interpolations with hermite and spline methods. The GRB contours correspond to the $68.3\%$ C.L. (with best fit at the black dot) for each calibration. The elipses correspond to $68.3\%, 95.4\%$ and $99.73\%$ C.L. constraints from the SDSS SNe Ia sample.}
\label{fig3}
\end{figure*}

\begin{table*} 
\begin{center}
\caption{Limits to $\Omega_{m0}$ and $h$ obtained from Figs. \ref{fig1} and \ref{fig2}.}
\begin{tabular}{r@{~~}c@{~~}c@{~~}c@{~~}c@{~~}c@{~~}c@{~~}}
\hline \hline \\
\multicolumn{1}{c}{ } & \multicolumn{3}{c}{$\Lambda(t)$CDM } & \multicolumn{3}{c}{$\Lambda$CDM } \\
\multicolumn{1}{c}{Test}&
\multicolumn{1}{c}{h}&
\multicolumn{1}{c}{$\Omega_{m0}$}&
\multicolumn{1}{c}{$\chi^2_{min}/\nu$}&
\multicolumn{1}{c}{h}&
\multicolumn{1}{c}{$\Omega_{m0}$ }&
\multicolumn{1}{c}{$\chi^2_{min}/\nu$}\\ \hline \\
Union2 (SALT2)......&0.697&0.338&0.978&0.700&0.269& 0.975 \\
NPZ......&0.728&0.463&0.868&0.814& 0.219&0.868\\
Hymnium......&0.708&0.422&0.405&0.776&0.206&0.405\\
MB......&0.757&0.440&0.630&0.834&0.215&0.631\\
\hline \hline
\end{tabular}
\end{center}
\end{table*}

With the new GRB samples of the last section we obtain the results shown in Fig. \ref{fig3} for the $\Lambda$(t)CDM (left panel) and $\Lambda$CDM (right panel) models. The $68.3\%$ C.L. (with best fit at the black dot) are shown for each calibration. Solid (dashed) open contours correspond to the cubic Hermite (spline) interpolation method. Table 3 shows the results corresponding to Fig. \ref{fig3}. Again, the large dispersion does not allow a definite conclusion, except that both models are in accordance with data within $1\sigma$ level. However, a look at the best-fit points is suggestive, specially if combined with the better posed results discussed in the Introduction. In the $\Lambda$CDM case the matter density parameter is in agreement with that resulted from the LSS analysis and in conflict with the larger value obtained with SDSS alone. If we consider instead the scenario with particle creation, we have again a very good concordance between the GRB and SDSS predictions for the present matter density, with both, Hermite and spline methods. If we also take into consideration the results for the Hubble constant, the concordance is particularly good with the Hermite interpolation. In none of the models the best-fit values for the matter density change when $h$ is marginalized.

\begin{table*} 
\begin{center}
\caption{Limits to $\Omega_{m0}$ and $h$ obtained from Fig. \ref{fig3}.}
\begin{tabular}{r@{~~}c@{~~}c@{~~}c@{~~}c@{~~}c@{~~}c@{~~}}
\hline \hline \\
\multicolumn{1}{c}{ } & \multicolumn{3}{c}{$\Lambda(t)$CDM } & \multicolumn{3}{c}{$\Lambda$CDM } \\
\multicolumn{1}{c}{Test}&
\multicolumn{1}{c}{h}&
\multicolumn{1}{c}{$\Omega_{m0}$}&
\multicolumn{1}{c}{$\chi^2_{min}/\nu$}&
\multicolumn{1}{c}{h}&
\multicolumn{1}{c}{$\Omega_{m0}$ }&
\multicolumn{1}{c}{$\chi^2_{min}/\nu$}\\ \hline \\
SDSS......&0.634&0.484&0.841&0.639&0.400&0.839  \\
GRB (Hermite)......&0.635&0.405&0.495&0.700&0.192&0.496\\
GRB (Spline)......&0.713&0.427&0.606&0.785&0.206&0.607\\
\hline \hline
\end{tabular}
\end{center}
\end{table*}

\begin{figure*}
\vspace{.2in}
\centerline{
\psfig{figure=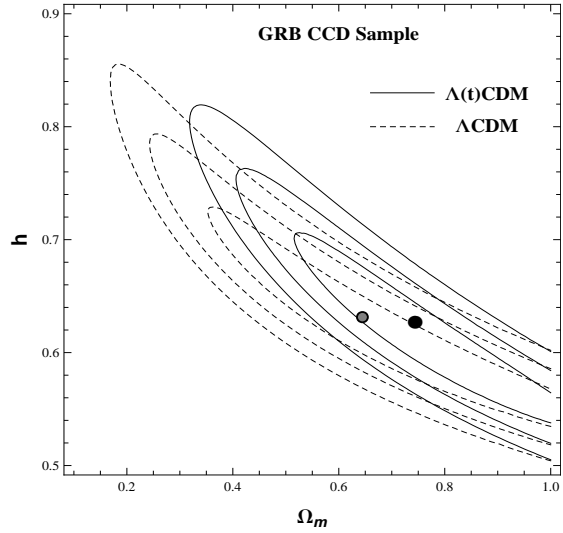,width=2.9truein,height=2.75truein}}
\caption{Contraints on the free parameters of the $\Lambda$(t)CDM (solid lines) and the $\Lambda$CDM (dashed lines) models. In this plot we have considered the CCD sample. The GRB contours correspond to $68.3\%, 95.4\%$ and $99.73\%$ C.L. (with best fit at the gray-filled and black dots, respectively).}
\label{fig4}
\end{figure*}

\begin{figure*}
\vspace{.2in}
\centerline{
\psfig{figure=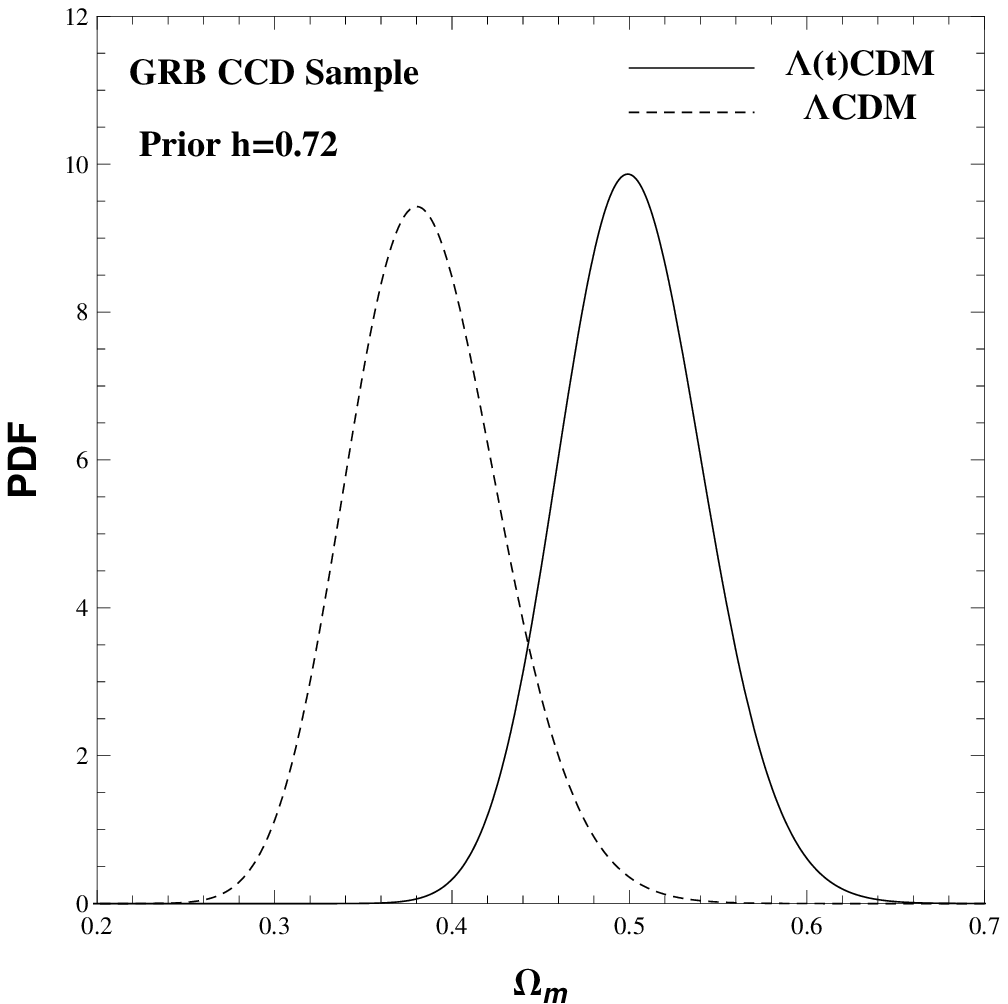,width=2.9truein,height=2.75truein}}
\caption{Probability distribution function for the matter density parameter $\Omega_{m}$ in the $\Lambda$(t)CDM (solid line) and the $\Lambda$CDM (dashed line) models. In this plot we have considered the CCD sample with prior $h = 0.72 \pm 0.08$.}
\label{fig5}
\end{figure*}

In our numerical analysis we have also included a distinct sample compiled by \cite{GRB2} (we will call it CCD sample from now on). This sample, being made of $69$ GRBs in the range $0.17<z<6.6$, is very particular because it makes use of a ``local regression calibration'' method (see \citep{GRB2} for details). The results are shown in Fig. \ref{fig4}. The high $\Omega_m$ values obtained here are in complete disagreement with the expected ones. This points out that the method and results of \cite{GRB2} are strongly dependent on the adopted priors. This is clear when we use, for the Hubble constant, the (rather conservative) prior $H_0 = 72 \pm 8$ km/s.Mpc \citep{Freedman}. With this prior, the probability distribution function for the matter density parameter $\Omega_{m}$ is given in Fig. \ref{fig5}, now in better accordance with our previous results.

However, \cite{GRB2} provides an interesting discussion about future contraints obtained with forthcoming GRB data, as expected by current missions like {\it Swift} and {\it Fermi}. Such analysis includes the computation of the Fisher matrix, leading to a prediction about the future error in estimating cosmological parameters. Concerning the comparison between $\Lambda(t)$CDM and $\Lambda$CDM models, the most important parameter is $\Omega_m$. From the results of \cite{GRB2}, the complementarity of future GRB and SN samples is able to reduce the uncertainty on the parameter $\Omega_m$ to $\sigma(\Omega_m)=0.019$ (for the $\Lambda$CDM model). With such an uncertainty level, $\delta\Omega_m \approx 0.02$, we will be surely able to place strong constraints on the dividing line between $\Lambda(t)$ and $\Lambda$ cosmologies.

In conclusion, we can see that the creation of particles from vacuum, something allowed by and generally expected from quantum field theories in curved spacetimes, is not only in agreement with current cosmological observations but is indeed favored by them when model-independent datasets are used. In the present model, where pressureless dark matter is produced with a constant rate, the vacuum density scales linearly with $H$ in order to conserve the total energy. Interestingly enough, this time-dependence for $\Lambda$ is precisely that derived when we estimate the renormalized density of the QCD vacuum condensate in the FLRW spacetime (see, for instance, \citep{PLB} and references therein). In this case the creation rate is $\Gamma \approx m^3$, where $m \approx 150$ MeV is the energy scale of the QCD chiral phase transition. This value for $\Gamma$ leads to a remarkable coincidence between $\Lambda$ and the observed dark energy density. Furthermore, as discussed in \cite{PLB}, the energy flux from vacuum to the matter sector also alleviates the cosmic coincidence problem. We think that all these theoretical and observational positive results are enough reasons to consider that scenario and to test it against a wider set of observations.

\section*{Acknowledgements}
This work was supported by CNPq (Brazil). HV acknowledges support from the DFG within the Research Training Group 1620 ``Models of Gravity''. AM acknowledges financial support from Conacyt-M\'exico, through a PhD grant. We are thankful to C. Pigozzo for his help with the SDSS statistical analysis.

\label{lastpage}

\end{document}